\documentclass{aa}
\usepackage{graphicx}
\usepackage{txfonts}
\newcommand{\uJy}{$\mu$Jy\,}
\newcommand{\lradio}{$L_{\rm 150\,MHz}$\,}
\newcommand{\mstar}{$M_{*}$\,}
\newcommand{\mr}{$M_{*}/R_{90}$\,}
\newcommand{\frl}{$F_{\rm RLAGN}$\,}

\defcitealias{Sabater19}{S19}
\defcitealias{Barisic19}{B19}
\usepackage[normalem]{ulem}

\begin{document}

   \title{Link between radio-loud AGNs and host-galaxy shape}

   \author{X.~C. Zheng\inst{1}
   \and H.~J.~A.~R\"{o}ttgering\inst{1} \and P.~N.~Best\inst{2}
   \and A.~van der Wel\inst{3,4}
   \and M.~J.~Hardcastle\inst{5}
   \and W.~L.~Williams\inst{1}
   \and M.~Bonato\inst{6,7}
   \and I.~Prandoni\inst{8}
   \and D.~J.~B.~Smith\inst{3} 
   \and S.~K.~Leslie\inst{1} 
          }
   \institute{Leiden Observatory, Leiden University, PO Box 9513, NL-2300 RA Leiden, the Netherlands\\
              \email{zheng@strw.leidenuniv.nl}
              \and
              SUPA, Institute for Astronomy, Royal Observatory, Blackford Hill, Edinburgh, EH9 3HJ, UK
              \and
              Sterrenkundig Observatorium, Department of Physics and Astronomy, Ghent University, Belgium
              \and
              Max-Planck Institut f\"{u}r Astronomie, K\"{o}nigstuhl 17, D-69117 Heidelberg, Germany
              \and
              Centre for Astrophysics Research, University of Hertfordshire, College Lane, Hatfield AL10 9AB, UK
              \and
              INAF - Istituto di Radioastronomia and Italian ALMA Regional Centre, Via Gobetti 101, I-40129, Bologna, Italy
              \and
              INAF - Osservatorio Astronomico di Padova, Vicolo dell'Osservatorio 5, I-35122, Padova, Italy
              \and
              INAF - Istituto di Radioastronomia, Via P. Gobetti 101, 40129 Italy
             }

   \date{}

 
  \abstract{
  The morphology of quiescent galaxies has been found to be correlated with the activity of their central super massive black hole.
In this work, we use data from the first data release of the LOFAR Two$-$Metre Sky Survey (LoTSS DR1) and the Sloan Digital Sky Survey Data Release 7 (SDSS DR7) to select more than 15 000 quiescent galaxies at $z<0.3$ to investigate the connection between radio-loud active galactic nuclei (RLAGNs) and the morphology of their host galaxy.
Taking advantage of the depth of LoTSS, we find that the fraction of RLAGNs with $L_{\rm 150\,MHz}>10^{21}\rm\,W\,Hz^{-1}$ at fixed stellar mass, velocity dispersion, or surface mass density does not depend on the galaxy projected axis ratio ($q$). 
However, the high-power ($L_{\rm 150\,MHz}>10^{23}\rm\,W\,Hz^{-1}$) RLAGNs are more likely to be found in massive, round galaxies, while the low- and intermediate-power ($L_{\rm 150\,MHz}\leq10^{23}\rm\,W\,Hz^{-1}$) RLAGNs have similar distributions of $q$ to non-RLAGN galaxies.
We argue that our results support the picture that high-power RLAGNs are more easily triggered in galaxies with a merger-rich history, while low-power RLAGNs can be triggered in galaxies growing mainly via secular processes.
Our work also supports the idea that the low-luminosity RLAGN may be sufficient for maintenance-mode feedback in low-mass quiescent galaxies with disc-like morphology, which is based on a simple extrapolation from the observed energy balance between cooling and RLAGN-induced cavities in massive clusters.
We find no significant difference between the $q$ distributions of RLAGNs likely to be found in clusters and those likely not found in clusters after controlling the radio luminosity and stellar mass of the two samples, indicating that the environment does not significantly influence the morphology--RLAGN correlation. 
  }


   \maketitle
%
\section{Introduction}

It is well established that there is a super massive black hole (SMBH) in the centre of almost all massive galaxies  \citep{Magorrian98,Kormendy04,Ho08} and strong correlations are found between the properties of the SMBHs and their hosts  \citep[e.g.][]{Gebhardt00,Greene06}, suggesting that the formation of galaxies is connected to the central SMBH activity.

Active central SMBHs, that is, active galactic nuclei (AGNs), emit huge amounts of energy in the form of either electromagnetic radiation or powerful jets from accretion.
Different Eddington-scaled accretion rates and the relative importance of the two kinds of output divide the AGNs into two categories: radiative- and jet-mode AGNs   \citep{Antonucci12,Best12,Heckman14}. 
The radiative-mode AGNs are generally less massive black holes with higher accretion rates and are usually found in star-forming (SF) galaxies, while the jet-mode AGNs are more likely to be in the centres of more massive, early-type galaxies.
The two types are associated with high-excitation radio galaxies (HERGs) and low-excitation radio galaxies (LERGs) \citep{Hardcastle07}.
Their powerful outflows and jets are believed to {heat the interstellar material and intercluster medium and prevent further star forming in their host galaxies}  \citep{McNamara07,Fabian12,Heckman14}. 

This AGN dichotomy is supported by studies based on large-scale radio surveys such as the NRAO VLA Sky Survey  \citep[NVSS;][]{Condon98}, Faint Images of the Radio Sky at Twenty centimetres  \citep[FIRST;][]{Becker95}, and LoTSS  \citep{Shimwell17}, in combination with some large-scale spectroscopic surveys, such as for example the Two-degree Field Galaxy Redshift Survey  \citep[2dFGRS;][]{Colless01}, the SDSS  \citep{York00}.
The local luminosity function of RLAGNs has been accurately measured in previous works such as \citet{Sadler02,Best05a,Hardcastle16} and \citet[hereafter \citetalias{Sabater19}]{Sabater19}. 
 \citet{Best05b} found the incident rate of RLAGNs in massive galaxies strongly depends on the stellar mass and black hole mass. 
More remarkably, using LoTSS and SDSS data, \citetalias{Sabater19} confirmed that the massive galaxies (i.e. \mstar$>10^{11}\,M_{\odot}$) are always switched on with $L_{150\rm\,MHz}\geq10^{21}\rm\,W\,Hz^{-1}$.
Using these results, the average heating rate due to recurrent jet activity can be estimated and compared to the radiative cooling rate of hot gas in the halos derived from X-ray observations  \citep{Best06,McNamara07,Hardcastle18}. 
It is then concluded that AGN radio jet power can supply enough energy to prevent gas cooling from the halos and quench star formation, thereby keeping galaxies `red and dead'.

{Despite plenty of evidence showing coevolution between AGNs and host galaxies, many details related to the build-up of the connection of SMBHs and galaxies are not yet clear.}
In particular, the basic question of how an RLAGN is triggered is still under investigation.
A commonly accepted idea is that black hole spin plays an important role  \citep{Wilson95,Hughes03,Fanidakis11}.
The Blandford--Znajek (B--Z) mechanism  \citep{Blandford77}, the most popular analytical jet-launching model, describes how a rotating black hole in a strong magnetic field converts the rotation energy to produce a relativistic jet.
The next part of the question relates to how a SMBH is spun up.
Such spinning up can be achieved by either major mergers of two massive black holes or a series of accretion events induced by minor mergers.
The two different spin-up paths are thought to be responsible for the SMBH spin in the most massive elliptical galaxies and in the less massive discy ones, respectively \citep{Sikora07,Fanidakis11}.
Therefore, it is important to study the correlation between RLAGNs and the morphology of their hosts. 

The correlation between RLAGNs and host morphology has been studied by many authors since the 1970s.
Observation of local powerful radio sources reveals that they are predominantly hosted by massive elliptical galaxies  \citep[e.g.][]{Condon78,Balick82,Smith86,Dunlop03,Best05b}.
Recently, \citet[hereafter \citetalias{Barisic19},]{Barisic19} studied the host-galaxy shape of the RLAGNs in the NVSS  \citep{Condon98} 1.4 GHz survey and the FIRST  \citep{Becker95} 1.4 GHz survey with SDSS data,
and confirmed with a large early-type galaxy sample that the incidence rate of RLAGNs with $L_{1.4\rm\,GHz}>10^{23} \rm\,W\,Hz^{-1}$ increases with the optical axis ratio of the 
host galaxy at fixed stellar mass and velocity dispersion.
%
{These studies focusing on the massive and luminous radio sources imply that powerful radio jets tend to be produced in galaxies with a merger-rich history.}

However, are the low- and intermediate-power radio sources also triggered by mergers?
In contrast to the powerful radio galaxies, a large part of low- and intermediate-power sources are late-type and/or show no merger signatures from high-resolution optical observations  \citep{Sadler14,Tadhunter16,Pierce19}.
A disc component can often be seen in their optical images via two-dimensional decomposition  \citep{Tadhunter16,Wang16,Wang19}.
It has therefore been suggested that these less powerful radio sources are more likely to be associated with secular processes (e.g. disc instabilities or bar-related processes) instead of mergers.

In this work, we extend the work of \citetalias{Barisic19} making use of the high-quality LOFAR data to study the morphology--radio power relation in a large sample with a low radio power limit to see whether or not there is any difference between the triggering of low- and high-power RLAGNs.
The article is structured as follows.
Section \ref{sec:data} describes our data and sample selection.
In Sect. \ref{sec:analysis}, we show how we deal with possible biases and analyse the connection between RLAGN fraction and galaxy projected axis ratio.
A discussion of the physical implications is presented in Sect. \ref{sec:discussion}.
Lastly, we provide a summary of our results in Sect. \ref{sec:conclusion}.
A standard cosmology with $H_0=70\rm\,km\,s^{-1}\,Mpc^{-1}$, $\Omega_{M}=0.3$, $\Omega_{\Lambda}=0.7$ is used throughout the paper.
\section{Data and sample selection}\label{sec:data}

Our radio sample is based on LoTSS  \citep{Shimwell17}, a high-resolution low-frequency (120 to 168 MHz) survey aiming to cover the whole northern hemisphere. 
The first data release of LoTSS  \citep[LoTSS DR1;][]{Shimwell19} contains over 300 000 sources with S/N$>$5, covering a region of 424 square degrees centred on the Hobby-Eberly Telescope Dark Energy Experiment  \citep[HETDEX;][]{Hill08} Spring Field.
With a median rms of $S_{144\,\rm MHz} = 71 $\uJy beam$^{-1}$, the survey achieves a point-source completeness of 90\%\ at a flux density of 0.45 mJy.
The 6 arcsec angular resolution and 0.2 arcsec positional accuracy allows for reliable optical and/or infrared (IR) counterparts.
 \citet{Williams19} performed careful source associations and optical and IR identification for the LoTSS radio sources with data from the Panoramic Survey Telescope and Rapid Response System  \citep[Pan-STARRS; ][]{Kaiser02} and the Wide-field Infrared Survey Explorer  \citep[WISE; ][]{Wright10} surveys.
As a result, 73\%\ of the DR1 radio sources have a reliable optical and/or mid-infrared (MIR) counterpart listed in the released value-added catalogue online \footnote{http://www.lofar-surveys.org}.

In a deep radio survey, the synchrotron emission from shocks due to supernovae in SF galaxies instead of central black hole activities can cause severe contamination in the final RLAGN sample.
To study the RLAGNs in the LoTSS sample, \citetalias{Sabater19} cross-matched LoTSS DR1 and the SDSS DR7 main galaxy sample \citep{Strauss02} and carefully separated RLAGNs from SF galaxies based on radio luminosity, WISE colours, and spectroscopic information such as emission line luminosity and $D_{4000}$ from SDSS, in particular the MPA-JHU value-added catalogue  \citep{Brinchmann04}.
SDSS DR7 provides 33324 sources for detailed photometric data, including flux, axis ratio, and radii in {\it ugriz} bands.
Of the 33324 galaxies, 10564 have a radio counterpart with a reliable classification.
In addition, we make use of the SED fitting results from  \citet{Chang15}, who combined SDSS and WISE photometry to make a comprehensive catalogue for galaxies in SDSS.
This catalogue provides useful information such as stellar mass and star-formation rate for 32663 galaxies and 10357 radio sources.
\begin{figure*}
 \centering
 \includegraphics[width=\linewidth]{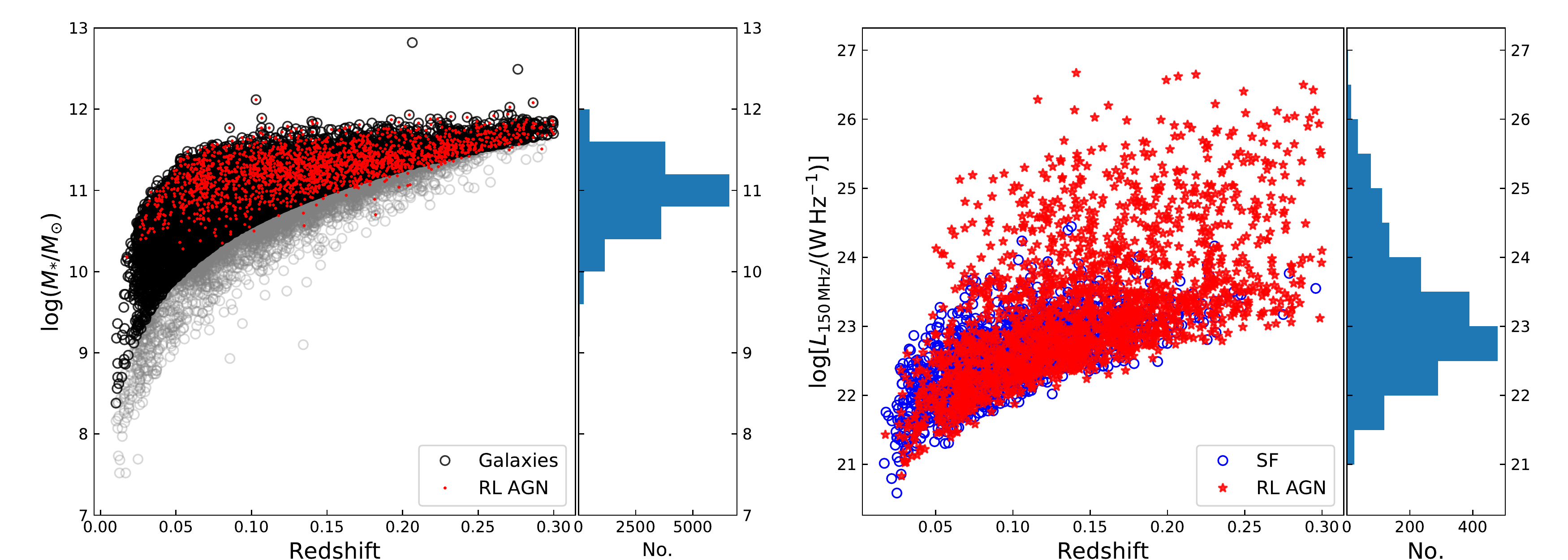}
 \caption{(Left) Stellar mass vs. redshift for the galaxies sample. 
 The black circles are galaxies in our final sample, while the RLAGN hosts are denoted with red points. 
 The grey circles are galaxies outside the mass completeness limit. 
 Radio-loud AGNs in this part are not used in the analysis.
 The histogram shows the distribution of the final galaxy sample. 
 (Right) Radio luminosity vs. redshift for the radio sources in the final sample. 
 The red stars denote RLAGNs while the blue circles are SF galaxies from radio analysis in  \citetalias{Sabater19}. 
 The histogram on the right represents the $L_{150\rm\, MHz}$ distribution of the RLAGNs in the final sample.}
 \label{fig:mldist}
\end{figure*}

For simplicity, apart from the SF/AGN separation from  \citetalias{Sabater19}, we also constrain our analysis to only the colour-based quiescent galaxies defined by the  \citet{Chang15} rest-frame colour--colour criteria: $u-r>1.6\times(r-z)+1.1$.
Moreover, we also restrict our analysis to galaxies above the SDSS DR7 mass completeness limit, ${\rm log}(M_{\rm limit}/M_{\odot})=10.6+2.28\,{\rm log}(z/0.1)$ \citep{Chang15}. 
This limit is also adopted to avoid biasing results by some low-mass high-luminosity sources. 
The final sample in our following analysis therefore contains 15934 colour-based quiescent galaxies, of which 3661 galaxies are associated with a radio source and 1912 galaxies are hosts of RLAGNs with \lradio $\ge10^{21}\,\rm W\,Hz^{-1}$.
{It should be noted that although the colour--colour { criterion rules out the selection of} non-quiescent galaxies, which are more likely to have a disc-like morphology, {  only 38 of the non-quiescent galaxies with RLAGN} are above the mass limit.
Therefore, the colour--colour criteria is not likely to  significantly influence the morphology distribution of the final sample.}

{We present the stellar mass distribution of the galaxies and the radio luminosity distribution of the RLAGNs and SF galaxies in Fig. \ref{fig:mldist}.
Similar to  \citetalias{Sabater19}, all the sources in our sample are within redshift 0.01 to 0.3.
{Within this redshift range, we confirm that the axis ratio distributions of galaxies at fixed \mstar do not show significant variation {with redshift (see Appendix \ref{app:qz})}, and therefore the bias of axis ratio measurement{  due to the small angular sizes of the galaxies at high redshifts} and the evolution of morphology distribution of galaxies are not significant in this work.}
According to the SED fitting, the stellar masses within our sample range from $10^8 \rm\, M_{\odot}$ to $10^{12} \rm\, M_{\odot}$.
{The stellar masses are based on results from  \citet{Chang15} if available; otherwise a MPA-JHU catalogue result is adopted.}

We also make use of the results from  \citet{Croston19} to investigate the environment of our sample.
 \citet{Croston19} matched the RLAGNs in LoTSS DR1 with two SDSS cluster catalogues  \citep{Wen12,Rykoff14} and calculated the cluster association probability for each RLAGN with $0.08<z<0.4$.
There are 1606 RLAGNs in our sample with environment information in the  \citet{Croston19} catalogue.
Of these, 460  have a high probability (>80\%) of being associated with a cluster, while 1138 have a probability of less than 50\%.}

\section{Data analysis}\label{sec:analysis}

\subsection{RLAGN fraction versus axis ratio }\label{sec:RLF}

Firstly, we would like to see how the radio-loud fraction (\frl) changes with projected axis ratio.
The strong stellar mass dependence of both galaxy morphology and the \frl can be an important source of bias.
{More massive galaxy samples have a larger fraction of elliptical galaxies, which have a greater chance of appearing `round', and therefore we would expect to see a larger fraction of round galaxies in a more massive sample.}

The \frl is also found to be strongly correlated with stellar mass  \citepalias[e.g.][]{Sabater19}.
Therefore, it would not be surprising to find that the \frl increases with projected axis ratio $q$ if the stellar mass is not taken into account.

\begin{figure*}
 \centering
 \includegraphics[width=0.95\linewidth]{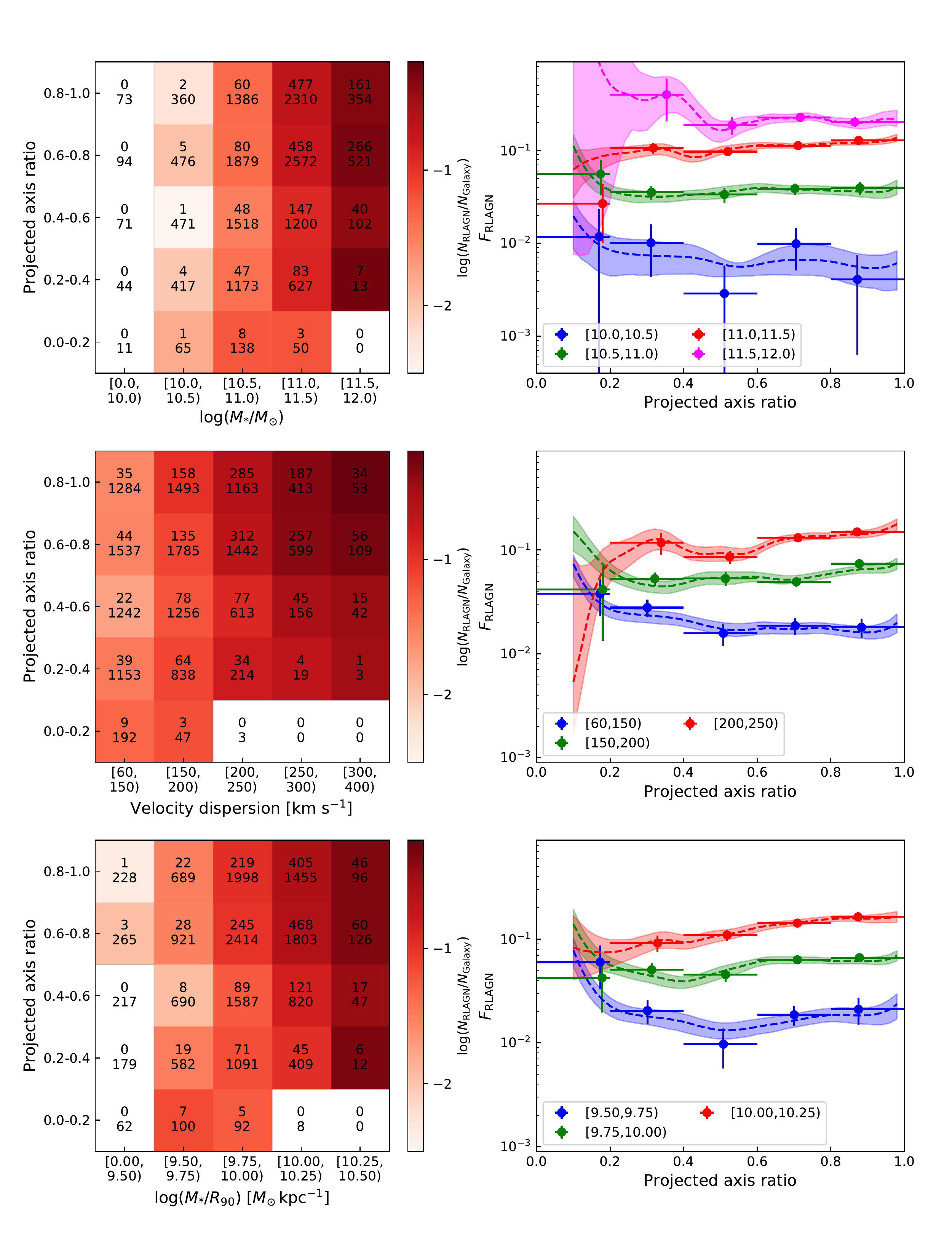}
 \caption{(Top left) Numbers of RLAGNs (top numbers) and galaxies (bottom numbers) in different groups of projected axis ratio in $r$ band and $M_*$.
 (Top right) Fraction of RLAGNs as a function of host galaxy projected axis ratio in different $M_{*}$ ranges.
 Results for different $M_{*}$ groups are colour coded.
 The ranges of $M_{*}$ are shown in the legend.
 The KDE-based (see text) results are displayed in dashed lines.
 The shaded areas represent the 1-$\sigma$ errors of the KDE results derived from 500 bootstraps.
 (Middle) Similar to the top two panels but grouped based on the stellar velocity dispersion $\sigma_{*}$ provided in the MPA-JHU catalogue \citep{Brinchmann04}.
 The results for the two largest $\sigma_{*}$ groups are not shown in the right panel because the results are affected by the noticeable larger \mstar at larger $q$ in these groups.
 (Bottom) Similar to the top two panels but grouped based on $M_{*}/R_{90}$.
 The results for the largest and smallest $M_{*}/R_{90}$ groups are not shown because of the possible bias and the small number of RLAGNs.
 }
 \label{fig:RLq_mass_hm}
\end{figure*}

To disentangle the \frl--$q$ correlation from the possible bias caused by the stellar mass, both the AGN and galaxy sample are grouped based on the stellar mass and $q$ in the $r$ band.
The grouping results are shown in the heat map in the top left panel of Fig. \ref{fig:RLq_mass_hm}.
We find that massive galaxies have a higher chance of hosting a RLAGN, which is in line with previous research  \citep[e.g.][]{Heckman14,Barisic19,Sabater19}.
However, there is no significant monotonic trend of RLAGN fraction along the $q$ axis, which is in stark contrast with the increasing trend shown in  \citetalias{Barisic19} using only the RLAGN with $L_{1.4 \rm GHz} > 10^{23} \,\rm W\,Hz^{-1}$.

{Because the LoTSS DR1 is not a volume-limited sample, the ratio of the numbers of RLAGNs and galaxies may not be an accurate indicator of the RLAGN fraction.
Here we use the definition in  \citet{Janssen12} and   \citet{Williams18}:
\begin{equation}
    F_{\rm RLAGN}^{M_{*},q}=(\sum_{i\in R_{M_{*},q}}\frac{1}{V_{i}})/(\sum_{j\in G_{M_{*},q}}\frac{1}{V_{j}}) \label{eq:frl}
,\end{equation}
where $R$ and $G$ are the RLAGNs and galaxies in a given group,
$V$ is the maximum accessible volume of a given source and is determined by the optical flux limit of SDSS main galaxy sample  \citep[14.5<$r$<17.77,][]{Strauss02} and the sky area distribution of noise in LoTSS DR1  \citep[see Fig. 12 in][]{Shimwell19}, as well as the redshift constraints in this work.
}

We plot the \frl as a function of $q$ in the right panel of Fig. \ref{fig:RLq_mass_hm}, where the galaxies are also divided into groups according to stellar mass.
None of the galaxy groups show a prominent increasing trend. 
For the highest-mass groups (${\rm log}(M_{*}/M_{\odot})>11.5$), the \frl is almost constant, which is not surprising as \citetalias{Sabater19}  found that the most massive galaxies are always switched on in LoTSS at 150MHz.
{However, the flat trend can also be seen in the lower mass groups.
This means that the triggering probability of a radio jet at fixed \mstar is independent of the galaxy axis ratio.
There is also a hint of a lowest point at $q\approx0.5$ in the lowest mass group but this is not statistically significant.
} 

\begin{figure*}
 \centering
 \includegraphics[width=\linewidth]{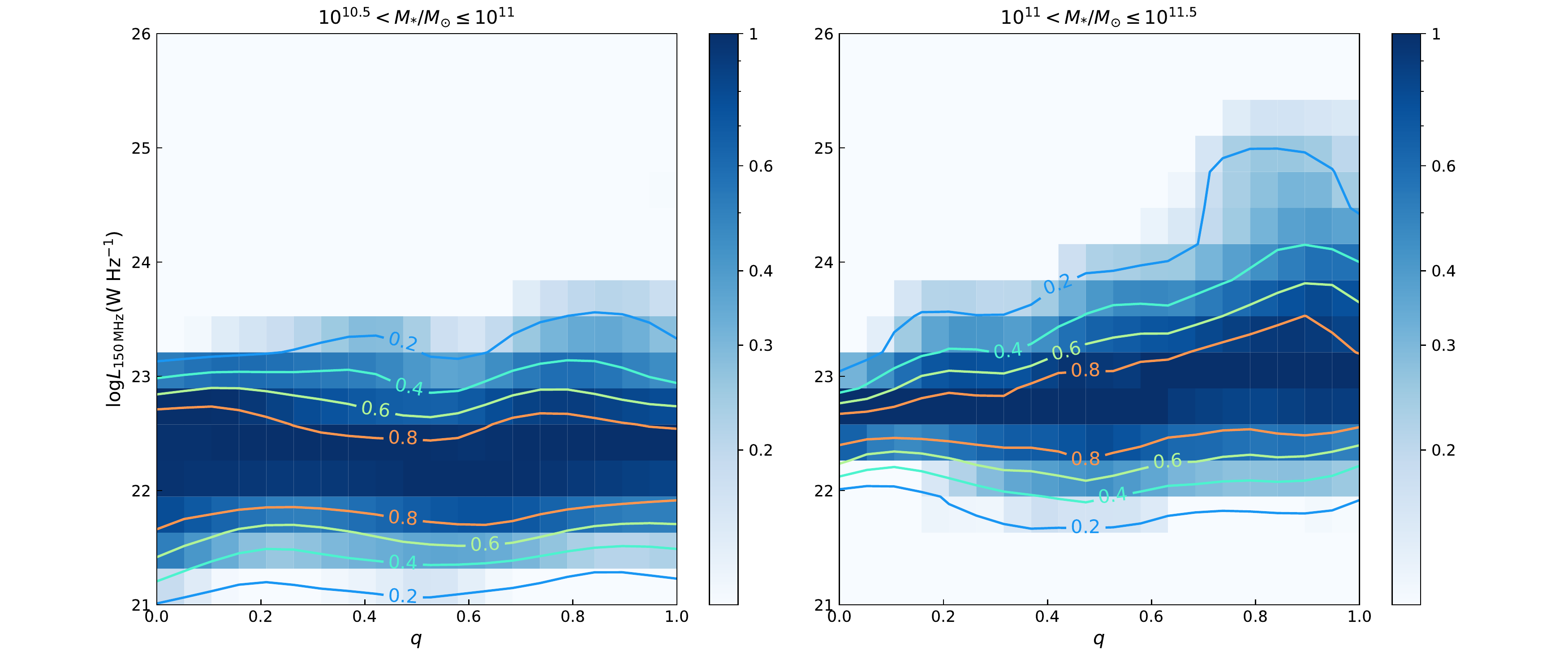}
 \caption{Normalized distribution of RLAGN at fixed \mstar in the \lradio--$q$ plane.
 Kernel-density estimation was performed to reduce the error from binning.
 The colour of each cell represents the source number relative to the maximum source number cell at fixed $q$.
 }
 \label{fig:Lvsq}
\end{figure*}
As well as grouping the sources according to stellar mass, we also group the sources based on stellar velocity dispersion $\sigma_{*}$ \footnote{Sources with $\sigma_{*}< 60\,\rm km\, s^{-1}$ are not considered because the $\sigma_{*}$ measurements are not reliable.}, {and the ratio of stellar mass and galaxy size $M_{*}/R_{90}$, which is also an indicator of velocity dispersion but is less affected by ordered rotational velocity.}
These two parameters represent the central black hole mass according to the well-known $M_{\rm BH}-\sigma_{*}$ relation  \citep[see ][for a review]{Kormendy13}.
The results are shown in the middle and bottom panels of Fig. \ref{fig:RLq_mass_hm}.

{The \frl -- $q$ relation in most of these groups exhibits similar features to that of the low-mass groups.
None of the groups display a strong correlation between \frl and $q$.
Only the group with $M_{*}/R_{90}\in[10,10.25)$ shows a weak increasing trend.}
It should be noted that we do not plot all groups in the right panels.
On the one hand, the groups with the largest $\sigma_{*}$ or $M_{*}/R_{90}$ have a noticeably higher average $M_{*}$ in rounder samples, which could lead to significant bias.
On the other hand, the smallest $M_{*}/R_{90}$ group is too small to lead to statistically significant results because of the lack of RLAGNs.

\begin{figure*}
 \centering
 \includegraphics[width=\linewidth]{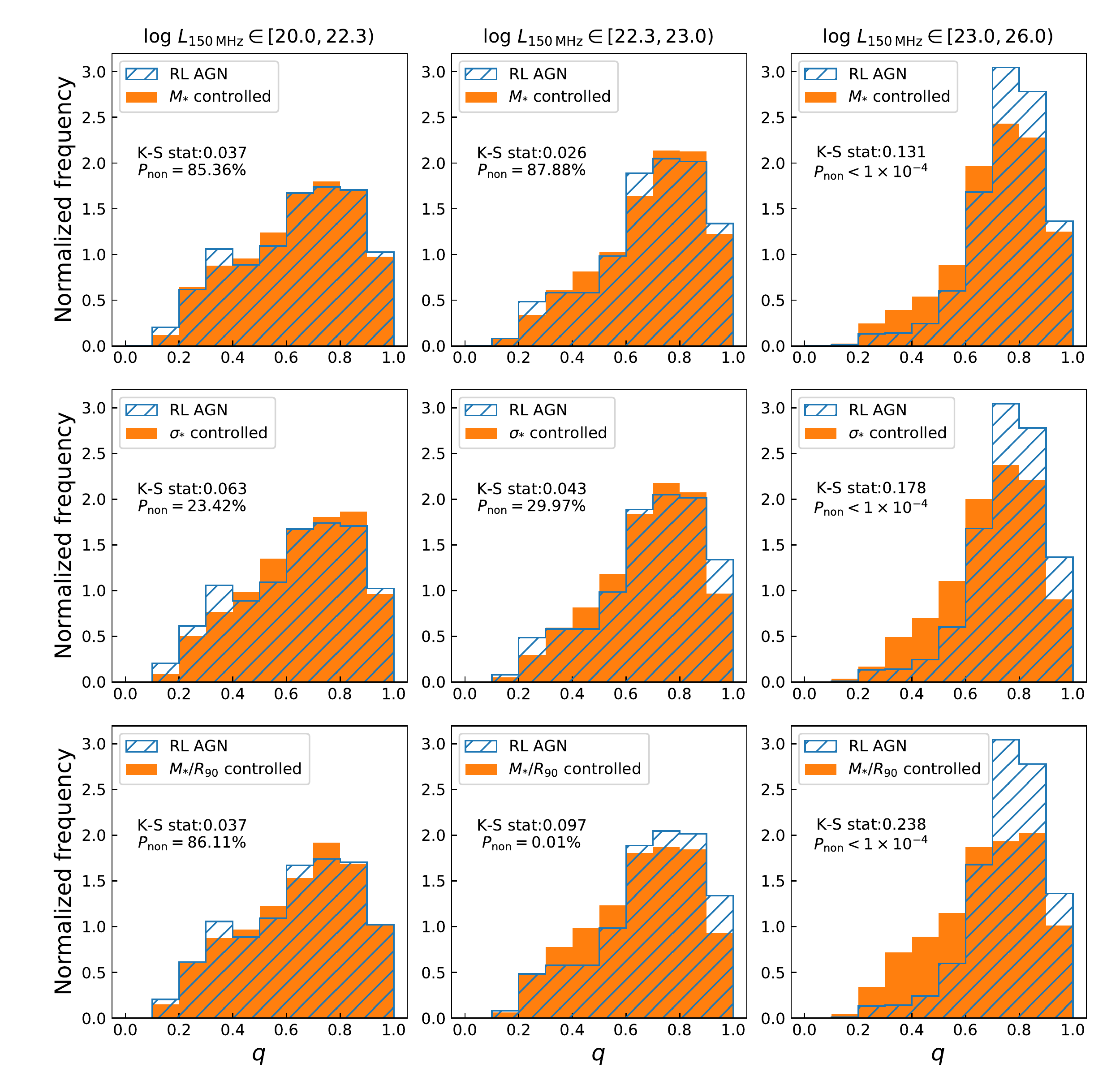}
 \caption{Normalized $q$ distributions of RLAGN hosts with different radio power and their control sample.
 From top to bottom: $q$ distributions of RLAGN hosts and \mstar, $\sigma_*$- and \mr- controlled sample.
 From left to right: $q$ distributions of low-, intermediate-, and high-power RLAGN hosts and their control sample. 
 The orange histograms show the distribution of `normal' galaxies, while the blue hatched histograms show the distribution of the RLAGN hosts.
 The $p$-values from K-S tests are listed in the panels.
 All the histograms are normalized for comparison.
 }
 \label{fig:RL_mass_control}
\end{figure*}
{We notice that there are some fluctuations in the \frl--$q$ relation especially in the lowest \mstar($\sigma_*$,\mr) groups. This is possibly the result of random binning effects, and therefore we also perform a kernel-density estimation (KDE) to reduce errors caused by binning.
We estimate the $q$ distributions of both RLAGN hosts and galaxies in each \mstar($\sigma_*$,\mr) group with Gaussian kernels using the method \texttt{gaussian\_kde} in \texttt{scipy.stats}  \citep{Virtanen19} with a rule-of-thumb bandwidth estimator based on Scott's rule  \citep{Scott92}.
We thus obtain $q$-smoothed \frl--$q$ correlations as the dashed lines in Fig. \ref{fig:RLq_mass_hm}.
The errors of the KDE results are estimated based on 500 bootstraps, as shown in the shaded areas in Fig. \ref{fig:RLq_mass_hm}.
The overall trends in all groups are consistent with the binning results but show weaker fluctuations, indicating that these fluctuations are due to random binning effects in small samples.

In conclusion, the \frl at fixed \mstar has only weak or no dependence on the galaxy axis ratio.
The {highly elongated} galaxies in our sample have a higher \frl than the sample in  \citetalias{Barisic19}.
This is likely due to the lower luminosity limit in our work, which is discussed in the following sections.}

\begin{figure*}
 \centering
 \includegraphics[width=\linewidth]{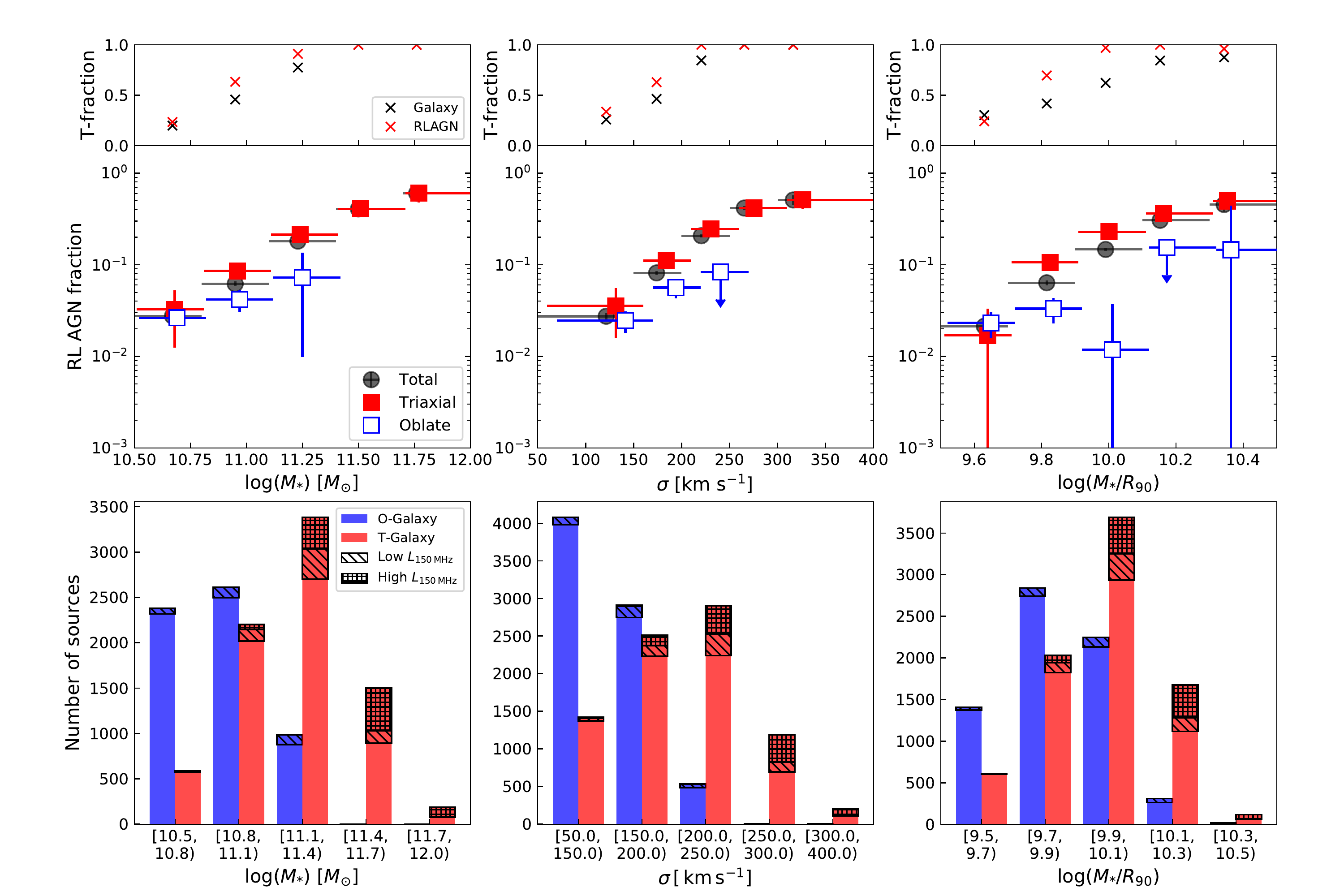}
 \caption{(Top) Triaxial source fractions as a function of \mstar, $\sigma_*$ , or \mr.
 The red crosses stand for the T-type fraction of the RLAGN sample and the black crosses are for the whole galaxy sample.
 (Middle) Radio-loud fraction of T/O-type sources as a function of \mstar, $\sigma_*$ , or \mr.
 Black points represent the whole sample, while the red(blue) rectangles are for the T(O)-type sample.
 Arrows mean that the corresponding points are upper limits.
 The horizontal error bars denote the bin sizes.
 The positions of the T(O)-type results are slightly shifted rightwards for clarity.
 (Bottom) Detailed decomposition results shown as bar charts.
 The red(blue) bars are the estimated numbers of the T(O)-type galaxies in different groups.
 The RLAGN results are overlaid as shaded areas on the galaxy results.
 Low- and high-power RLAGNs are marked with different hatching.
 }
 \label{fig:TO}
\end{figure*}

\subsection{Radio power dependence versus axis ratio}\label{sec:control}

{The great difference in the \frl--$q$ relation between our results and those of  \citetalias{Barisic19} implies that the shape of RLAGN hosts could be radio-power dependent.
To see how the shape of RLAGN hosts changes with radio power, we show the normalized RLAGN distribution at fixed \mstar in the \lradio--$q$ plane in Fig. \ref{fig:Lvsq}.
The value of each cell represents the number of RLAGNs with given radio luminosity and $q$ normalized to the number of RLAGNs in the most populated cell at given $q$, that is, the value of the cells in the darkest blue are always 1 and they mark the \lradio position of the source density peak at fixed $q$.
This normalisation is used to enhance the visualisation at very small and high $q$, where there are relatively small numbers of sources compared with the central part.

Figure \ref{fig:Lvsq} shows that at low \mstar ($<10^{11}\,M_{\odot}$), RLAGNs with different $q$ have similar \lradio distribution, and that
at high \mstar, only round ($q>0.6$) RLAGNs can reach a high radio power, which means that the radio power distribution of RLAGNs depends on the galaxy morphology.
Therefore, if we only inspect the RLAGNs brighter than $L_{1.4\rm\,GHz}=10^{23}\rm\,W\,Hz^{-1}$ (equivalent to \lradio$=10^{23.68}\rm\,W\,Hz^{-1}$ assuming a radio spectral index of 0.7) as in  \citetalias{Barisic19}, we can only find RLAGNs with a large axis ratio, which is the reason for the difference in conclusions between our work and that of  \citetalias{Barisic19}.

It is also important to investigate the difference between the shape of normal galaxies and RLAGNs with different radio power.
}
To avoid the bias from $M_{*}$ affecting our comparison, we construct mass-controlled samples based on the distribution of masses of RLAGN hosts.

Firstly, we divide the whole RLAGN sample into three subsamples with different radio luminosity ranges ($20\leq{\rm log}L_{150\,\rm MHz}<22.3$, $22.3\leq{\rm log}L_{150\,\rm MHz}<23$,
$23\leq{\rm log}L_{150\,\rm MHz}<26$).
The three subsamples represent the low-, intermediate-, and high-radio power sources in the following analysis.
{Secondly, we obtain the \mstar distributions of the RLAGN hosts by counting the numbers of galaxies in eight logarithmic equidistant bins ranging from $10^{10}$ to $10^{12}$ $M_{\odot}$ for each subsample, in which each mass bin contains between 0 and 400 RLAGNs.
The normal galaxies, defined as galaxies not hosting an RLAGN, are also grouped in the same \mstar bins and the number of galaxies in each bin ranges from 75 to 4000.
Once the RLAGN distribution is known, we randomly select 3000 (about three times the number of the largest RLAGN subsample to reduce random errors) normal galaxies but with the same \mstar distribution as the RLAGN sample \footnote{There may not be enough normal galaxies within the highest \mstar bins, and therefore some galaxies can be selected more than once.}.
These sources then become the \mstar-controlled sample of the associated RLAGN subsample.}
In this way, we remove the \mstar bias, allowing us to make a direct comparison between the shape of RLAGN hosts and normal galaxies.
Moreover, following the same procedures, we also construct $\sigma_{*}$- and \mr- controlled samples for comparison.
The normalized $q$ distributions of the RLAGN and control sample are all shown in Fig. \ref{fig:RL_mass_control}.

First of all, it is apparent from Fig. \ref{fig:RL_mass_control} that high-power RLAGN samples have a larger fraction of round galaxies than the control samples.
We perform a Kolmogorov-Smirnov (K-S) test on all groups to measure the significance of the difference.
In all the high-power groups, the null hypothesis that the two samples are drawn from the same distribution can be rejected with  $p$-values all smaller than $1\times10^{-4}$.
Therefore, high-power RLAGN hosts do tend to appear rounder than the `normal' galaxies.
This is in line with what has been revealed in previous studies, namely that very few powerful RLAGNs are found in geometrically flat early-type galaxies.

Interestingly, in low- and intermediate-power RLAGN samples, no significant  excess of round galaxies is found.
The distributions of the RLAGN samples appear similar to their control sample.
The K-S tests show that only the intermediate-power \mr-controlled group shows a significant difference between the RLAGN sample and normal galaxies, while in the other groups the null hypothesis cannot be rejected ($p$-value > 5 \%).
Moreover, although the intermediate-power \mr-controlled group shows significant distinction, the excess in the fraction of round sources for RLAGNs is not as large as in the high-power groups.
Apparently, low- and intermediate-power RLAGNs do not appear to have a rounder morphology on average than normal galaxies.

In conclusion, while \frl does not strongly depend on $q$, the resultant radio luminosity distribution does.
While high-power RLAGNs are more likely to be visually round, the low- and intermediate-power RLAGNs have a similar $q$ distribution to the normal galaxies.

\subsection{T/O decomposition}

\citet{Emsellem11} and  \citet{Chang13} have demonstrated that the $q$ distribution of early-type galaxies can be described by a composition of two distinct populations of galaxies, triaxial (T) and oblate (O) types.
{In such an analysis, it is assumed that the three-dimensional light profile of the entire galaxy can be approximated by an oblate shape in spite of the detailed structure (e.g. bulge and disc component).}
{The T-type galaxies have a higher chance of showing a round shape along the line of sight, while the O-type sources have a wide distribution of $q$.}
\citetalias{Barisic19} applied the decomposition results from  \citet{Chang13} and concluded that most of the RLAGN hosts with $L_{\rm1.4\,GHz}>10^{23}\,\rm W\,Hz^{-1}$ are T-type galaxies.
In this section, we follow their methodology and perform the same decomposition to investigate the T/O populations in our sample.
\begin{figure*}
 \centering
 \includegraphics[width=\linewidth]{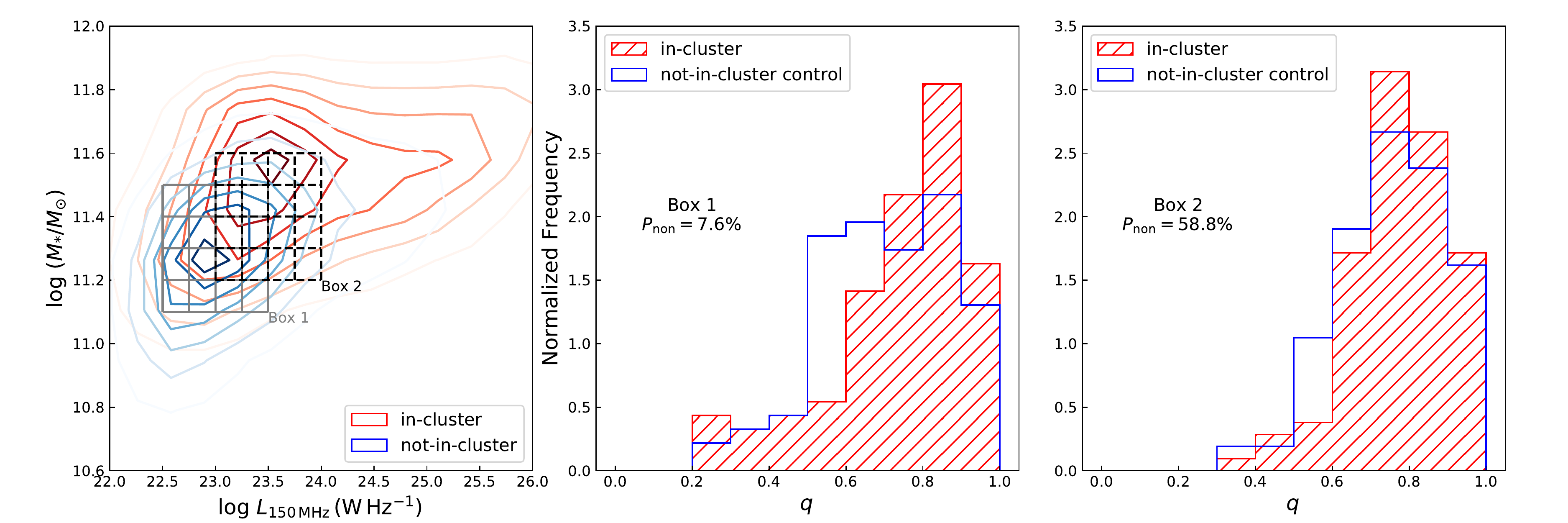}
 \caption{(Left) Distribution of RLAGNs with or without an associated cluster in \lradio--\mstar space.
 Red contours denote the distribution of in-cluster RLAGNs, while the blue contours denote the not-in-cluster RLAGNs (defined in Section \ref{sec:analysis:env}).
 The grey lines (`Box 1') and black dashed lines (`Box 2') mark the \lradio and \mstar constraints and bins used in the control sample analysis in Section \ref{sec:analysis:env}.
 The grey lines are for the low-power sample, while the black dashed lines are for the high-power sample.
 (Middle and Right) Example of the $q$ distributions of in-cluster RLAGNs (red shaded histograms) and the not-in-cluster control sample (blue histograms).
 The null hypothesis $p-$value from a K-S test (median value from 1000 tests) is noted.
 }
 \label{fig:env}
\end{figure*}

We first construct a simulated T/O sample containing 10000 sources with a distribution of the morphology parameters (e.g. triaxiality, axis ratio, and their dispersion) given by  \citet{Chang13}.
Randomly selected viewing angles are then assumed to calculate the expected normalized $q$ distribution for T- and O-type galaxies.
{These two model distributions are linearly combined with different weights to match the $q$ distribution of RLAGN hosts and galaxies in different \mstar ($\sigma_*$, \mr) groups. }
In this way we are able to estimate the fractions of T/O-type sources with the best-fit weights in each subsample and the RLAGN fractions of each population\footnote{Because it is not possible to locate the sources in each category, we cannot calculate the RLAGN fractions with Equation \ref{eq:frl} but the ratios of the expected numbers of RLAGNs and galaxies in a given bin. 
Nevertheless, we compared the RLAGN fractions based on the two methods and find that this will not affect the results significantly in this work.}.
It should be noted that the morphology parameters given in  \citet{Chang13} may not be a perfect model to fit the $q$ distribution of RLAGNs if RLAGNs have different sets of morphology parameter distribution intrinsically, but it gives a good indication of the relative importance of T- and O-type sources in each subsample and allows us to compare our results with those of  \citetalias{Barisic19}.
The results are presented in the top panels of Fig. \ref{fig:TO}.

First, we can see the T-type fraction strongly correlates with \mstar for both the RLAGN and galaxy samples.
In particular, the T-type fraction reaches 100\%\ for galaxies with $M_*>10^{11.5}\,M_{\odot}$.
Similar trends can also be seen in the $\sigma_*$ and \mr grouping results.
Furthermore, RLAGNs have a larger T-type fraction than the whole galaxy sample in almost all groups.
These findings are consistent with a massive elliptical galaxy having a higher chance of being a radio-loud source  \citep[e.g.][]{Matthews64,Wilson95} than a disc-like galaxy.

Second, both T- and O-type galaxies have an increasing \frl with \mstar.
Most importantly, although the \frl of T-type sources is larger than that of O-type sources, the two \frl are close to each other, especially in the low-\mstar part. 
This result is different from that of  \citetalias{Barisic19}, where O-type galaxies have only a negligible \frl compared with T-type sources.
{Considering that the main difference between our sample and that of  \citetalias{Barisic19} is the radio power, we suggest that this is the main reason for the discrepancy.}

To further investigate the importance of radio power, we divide the RLAGN sample into two subsamples according to whether a source \lradio is larger or smaller than $10^{23}\rm\,W\,Hz^{-1}$\footnote{This criterion is not exactly equivalent to the luminosity limit $L_{1.4\rm\,GHz}=10^{23}\rm\,W\,Hz^{-1}$ in  \citetalias{Barisic19}, because we have to ensure there are enough sources in each subsample for decomposition.}.
We then perform similar T/O decomposition to the two subsamples to estimate the number of low- and high-power RLAGNs in T/O-type galaxies.
The estimated numbers of different kinds of sources are presented in the bottom panels of Fig. \ref{fig:TO}.

As a result, we can see that while the RLAGNs in the T-type sample contain a considerable amount of both low- and high-power sources, those in the O-type sample are nearly all low-power AGNs.
This proves that low-power RLAGNs are the reason for the high \frl of O-type galaxies.

\subsection{Environment}\label{sec:analysis:env}

{
Environment also plays an important role in determining the properties of a galaxy.
On the one hand, RLAGNs are found to have a higher prevalence in cluster environments  \citep{Best05b,Tasse08,Sabater13}.
This tendency cannot be explained simply by the higher stellar mass of galaxies in more massive halos.
 \citet{Best05b} indicated that this clustering trend also depends on the emission-line luminosity. 
 \citet{Donoso10} and  \citet{Croston19} also found a strong correlation between clustering strength and radio power.
On the other hand,  \citet{Cappellari11b} showed that early-type galaxies with different stellar kinematic features, which are linked to galaxy morphology, have different correlations with the environment density.
In this case, the RLAGN--environment relation and the morphology--environment relation may result in a correlation between RLAGNs and galaxy morphology.
\begin{figure*}
 \centering
 \includegraphics[width=\linewidth]{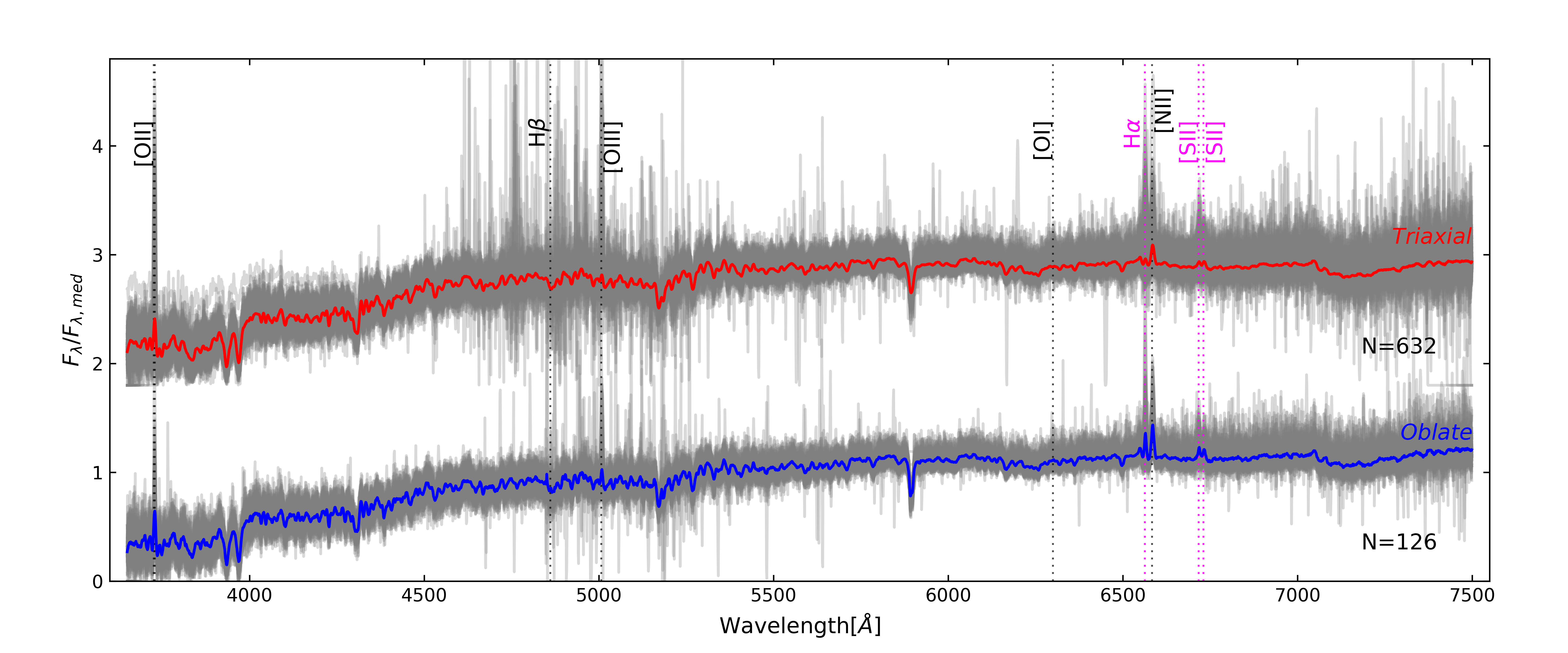}
 \caption{Stacking rest-frame spectra of T/O-type sources with $L_{\rm 150\,MHz}<10^{23}\,\rm W\,Hz^{-1}$.
 The red T-type spectrum is the average of all round ($q>0.6$) RLAGNs weighted by the possibility of each individual source being an intrinsic T-type.
 For clarity, the T-type stacking spectrum is shifted upwards.
 The blue O-type spectrum is derived in a similar way but using {highly elongated} ($q<0.4$) RLAGNs and weighted by the possibility of each source being an intrinsic O-type. 
 The grey shadows are the individual spectra of the sources used in the stacking.
 The emission lines needed for calculating the excitation index \citep{Buttiglione10} are marked in the figure.
 }
 \label{fig:spec}
\end{figure*}

{To study the role of environment in the RLAGN--morphology relation, we compared the shapes of 460 `in-cluster' RLAGNs and 1138 `not-in-cluster' RLAGNs using the catalogue of  \citet{Croston19}.
It should be noted that as  \citet{Croston19} made use of two cluster catalogues  \citep{Wen12,Rykoff14}, we define the in-cluster RLAGNs as sources with a cluster association probability of larger than 80\%\ in either of the two catalogues and the not-in-cluster RLAGNs as the sources with
a cluster association probability of less than 50\%\ in both catalogues.}
The distribution of these two samples in \lradio--\mstar space is presented in the left panel of Fig. \ref{fig:env}.
}

{
According to our comparison, the in-cluster RLAGNs are brighter and more massive than the not-in-cluster sample, which is consistent with the finding by  \citet{Croston19}.
Since the galaxy shape strongly depends on \lradio and \mstar, we cannot compare the shapes of the in-cluster and not-in-cluster RLAGNs directly.
Instead, we again use the control sample analysis to avoid biases from the different \lradio\ and \mstar distribution of the two samples.

We first require the RLAGNs used in the analysis to have a \lradio from $10^{22.5}\rm\,W\,Hz^{-1}$ to $10^{23.5}\rm\,W\,Hz^{-1}$ and a \mstar from $10^{11.1}\,M_{\odot}$ to $10^{11.5}\,M_{\odot}$.
These constraints are highlighted as `Box 1' in the left panel of Fig. \ref{fig:env}.
This choice ensures both samples have enough sources within the parameter range, and on the other hand, it allows us to study a sample with a large fraction of low-power (\lradio$<10^{23}\rm\,W\,Hz^{-1}$) RLAGNs.
As a result, the constraint leads to 92 in-cluster and 445 not-in-cluster RLAGNs.

After constraining the parameter space, we separate the space into 16 small regions with different \lradio and \mstar as shown in Box 1 of Fig. \ref{fig:env}.
We then construct the not-in-cluster control samples by resampling the not-in-cluster RLAGNs to force the source number of the control sample in each \lradio-\mstar region to match that of the in-cluster RLAGNs.
We use a K-S test to compare the $q$ distribution of the in-cluster and the control samples to assess the similarity of their shapes.
This resampling process is repeated 1000 times and the final null hypothesis $p$-value is based on the median of all K-S test results.

According to the K-S test result, we cannot reject the null hypothesis at a significance level of 0.05, which means that the shape difference between the two samples is not significant.
However, we notice that the $p$-value of 7.6\%\ is a marginal one.
In the middle panel of Fig. \ref{fig:env}, the in-cluster sample has a slightly larger fraction of high $q$ sources than the not-in-cluster sample.
Therefore, we suggest that it is still possible that environment has a weak influence on the shape of the RLAGN hosts, but this influence cannot be the main reason for the morphology--RLAGN relation.

{We also choose a higher power constraint as `Box 2' in Fig. \ref{fig:env} to perform the same analysis.
Box 2 requires the RLAGN to have a \lradio from $10^{23}\rm\,W\,Hz^{-1}$ to $10^{24}\rm\,W\,Hz^{-1}$  and a \mstar from $10^{11.2}\,M_{\odot}$ to $10^{11.6}\,M_{\odot}$.
This sample contains 105 in-cluster and 308 not-in-cluster RLAGNs.

As a result, the K-S test gives a much larger null hypothesis $p$-value (58.8\%). 
Therefore, the environment is not likely to have a direct impact on the connection between high-power RLAGNs and the morphology of their hosts.
}
}

\subsection{Accretion mode}\label{sec:analysis:edd}
{
From previous sections, we find that RLAGNs with different galaxy shapes have different distributions of \mstar and \lradio. 
At the same time, different RLAGNs have different accretion modes  \citep[see ][for more details]{Best12}.
Therefore, in this section we consider whether or not there is any difference between the accretion modes of the T/O-type sources.

Unfortunately, most of the sources in our sample do not have spectra of sufficient quality for a reliable high-/low-excitation classification.
Considering that an important distinction of these two accretion modes is the Eddington accretion rate, we simply separate the sources based on their Eddington ratio.
We adopt the $M_{\rm BH}-\sigma_*$ relation in  \citet{Bosch16} to calculate the black hole mass.
The radiative and jet mechanical luminosities are estimated based on the [OIII] 5007 luminosity  \citep{Heckman04} and $L_{\rm 1.4\,GHz}$  \citep{Cavagnolo10,Best12} as follows:
\begin{eqnarray}
    L_{\rm rad}/L_{\rm [OIII]}&\approx&3500 \\
    L_{\rm mech}&=&7.3\times10^{36}(L_{1.4\rm\,GHz}/10^{24}\rm\,W\, Hz^{-1})^{0.7} W \label{eq:radio-pow}
.\end{eqnarray}
The 1.4 GHz luminosity is derived from 150 MHz luminosity assuming a spectral index of 0.7.
The Eddington ratio is then derived by $\lambda_{\rm Edd} = [L_{\rm rad}+L_{\rm mech}]/L_{\rm Edd}$, where the Eddington luminosity is obtained from $L_{\rm Edd} = 1.3\times10^{31}\,M_{\rm BH}/M_{\odot}\,\rm W$.

As a result, we obtain the Eddington ratio for 1889 out of 1912 RLAGNs in our sample.
Only 69 sources have an Eddington ratio larger than $10^{-2}$, which is the adopted division line for quasar-mode and jet-mode AGNs  \citep{Best12}.
The resulting number of quasar-mode ($\lambda_{\rm Edd}>-2$) AGNs is much less than the expected number of either O-type ($\sim$300) or T-type sources ($\sim$1600).
Therefore, the accretion mode is not likely to have a simple link with the morphology.
{However, it should be noted that this result depends largely on the accuracy of the Eddington ratio estimation.
 \citet{Hardcastle19} discussed the relation between the radio luminosity and the jet kinetic power, and indicated that Equation \ref{eq:radio-pow} can be very unreliable especially at low luminosity.
Therefore, it is difficult to have a definite conclusion based on the current information.
}

We also use a spectrum-based method to inspect the differences in RLAGN properties.
We stack spectra for T/O-type galaxies using the archival spectra of our sample from the SDSS Science Archive Server  \citep{Smee13,Aguado19}.
Firstly, because the high-power ($L_{\rm 150\,MHz}\geq10^{23}\,\rm W\,Hz^{-1}$) RLAGNs are dominated by T-type sources, which may cause contamination to the O-type spectra, therefore we only use the low-power ($L_{\rm 150\,MHz}<10^{23}\,\rm W\,Hz^{-1}$) sources so as to reduce the number of the T-type galaxies.
Next, we estimate the T/O-type fraction of the sample and derive the possibility of each galaxy being a T/O-type source according to the projected axis ratio.
We then use \texttt{pysynphot} \citep{STScI13} to rebin the rest-frame spectrum of each source.
Finally, we weight each source spectrum (normalized by their median flux) by the possibility of being T/O-type and calculate the average spectrum for both types.
To reduce the contamination from each other type, we only use round galaxies ($q>0.6$) to extract T-type spectra and {highly elongated} galaxies ($q<0.4$) to extract O-type spectra. 
The spectra are presented in Fig. \ref{fig:spec}.

As we highlight in Fig. \ref{fig:spec}, the two spectra are almost identical except for the strength of the H$\alpha$ and [SII] emission lines.
Both spectra lack very strong emission lines, which implies that the two subsamples are both dominated by jet-mode AGNs.
This supports the idea that the morphology types do not favour an accretion-based dichotomy.
On the other hand, the stronger H$\alpha$ and [SII] lines in O-type spectrum may be a hint that O-type galaxies have more gas than T-type or they have different ionizing continua.
}

\section{Discussion}\label{sec:discussion}
We find that the low-power RLAGNs show a different connection with host-galaxy morphology compared with the high-power RLAGNs.
{However, a potential caveat regarding the results is the reliability of the AGN/SF classification at low radio luminosity.
The contamination of SF galaxies in the low-luminosity AGN sample may increase the fractions of {highly elongated} sources.
However, our work is based on the careful classifications of  \citetalias{Sabater19}.
These latter authors argued that the potential overall misclassification rate is less than 3\%.
Therefore, assuming that the misclassification rate is 6\%\ at \lradio$\leq 10^{23}\rm\,W\,Hz^{-1}$, we randomly changed the classifications of 6\%\ of the {highly elongated} ($q\leq0.5$) AGNs and round ($q>0.5$) SF galaxies with \lradio$\leq10^{23}\rm\,W\,Hz^{-1}$, and then repeated the analyses of the above sections to estimate the influence of the contamination.
Almost the same results were obtained, proving that the contamination of SF galaxies does not affect our conclusions.}

The relatively large fraction of oblate ({highly elongated}) sources in the low-power sample compared to the high-power sample implies that there could be two distinct populations of RLAGNs.
One of them tends to be in triaxial (round) galaxies and can be very powerful sources, whereas the other population favours oblate ({highly elongated}, or disc-like) galaxies.
This can naturally explain what we see in Fig. \ref{fig:RLq_mass_hm}.
The occurrence of two populations is probably related to different triggering mechanisms of radio jets.
{Since the two types of RLAGNs have similar accretion modes, a possible factor causing the difference is the spin-up paths of the SMBHs, assuming a B-Z jet launching model.
} 

The SMBH spin-up process can theoretically be linked to galaxy morphology.
 \citet{Chang13} suggested that the T/O early-type galaxies correspond to slow rotator (SR)/fast rotator (FR), a kinematically distinct dichotomy described in  \citet{Cappellari07,Emsellem07,Emsellem11}.
Massive SRs are more likely to be the end products of some extreme merger events, while the FRs could be `dead' spirals resulting from gaseous dissipative processes.
Theoretically, mergers can effectively give rise to SMBH rotation  \citep{Baker06,Fanidakis11}, resulting in a fast-spinning SMBH in a massive disturbed (possibly elliptical) galaxy.
As for the low-mass sources, which have not been through as many merger events, they can more easily keep or rebuild a disc component and the central black holes acquire their angular momentum mainly by accretion.
This picture is consistent with the findings of  \citet{Pierce19} and  \citet{Wang16,Wang19}, where intermediate- and low-mass (or luminosity) sources are found to be less morphologically disturbed.
{These long-term evolution processes may not directly trigger radio jets, but have an important influence on the triggering probability and properties of radio jets.}

{The association between T/O-type and SR/FR classification can be inferred from the relative fractions of each type. 
From Fig. \ref{fig:TO}, we find that the T-type fraction of the whole galaxy sample is $\sim20$\%\ at \mstar$<10^{10.8}\,M_{\odot}$, then rises rapidly to $\sim50$\%\ at $\sim10^{11}\,M_{\odot}$ and reaches $\sim100$\%\ after \mstar$>10^{11.4}\,M_{\odot}$.
This trend is consistent with the results of  \citet{Chang13}.
In comparison, integral-field spectroscopic (IFS) surveys such as ATLAS$^{\rm 3D}$  \citep{Cappellari11a}, MASSIVE \citep{Ma14}, the SAMI galaxy survey \citep{Bryant15}, and SDSS-IV MaNGA  \citep{Bundy15} have revealed that the fraction of SRs in early-type galaxies increase from less than 20\%\ at \mstar$<10^{11}\,M_{\odot}$ to about 90\%\ at \mstar$\sim 10^{12}\,M_{\odot}$  \citep{Cappellari11b,Sande17,Veale17a,Veale17b,Graham19}.
We can see the changing trends with \mstar for T-type galaxy fraction and SR fraction are very similar although the T-type fraction is slightly larger than the SR fraction at the same \mstar.
Therefore, we suggest that the association between T/O-type and SR/FR classification is reasonable.

In Sect. \ref{sec:analysis:env}, we investigate the $q$ distribution of the `in-cluster' and `not-in-cluster' RLAGNs, and find they are not significantly different, especially for the high-power RLAGNs.
It should be noted that this result does not conflict with the finding of  \citet{Croston19}, where they found large-scale environment is influential in driving AGN jet activity, because in Fig. \ref{fig:env} we do see the in-cluster sample has a larger fraction of high \lradio sources than the not-in-cluster sample at fixed \mstar.

{If the spin is the key reason causing different radio power distributions for galaxies with different shape, the morphological similarity can be interpreted in the following way: although a higher rate of merging events in denser environments leads to a larger fraction of high-spin SMBHs and therefore high-power RLAGNs, {RLAGNs with similar radio power and host stellar mass in different environments may have similar spin-up history.
%
In other words, the relative importance of the secular processes and mergers in the spin-up history of a RLAGN with a given radio luminosity and stellar mass is more likely to be independent of the current environment.
}
However, due to a lack of information on the spin of black holes, we are not yet able to confirm the role of spin in morphology--RLAGN relation directly.

{This work can also improve our understanding of the role of the maintenance-mode feedback.
It is suggested that the maintenance-mode feedback works when the cooling rate of the hot atmosphere balances with the time-average energy output from RLAGNs, which is determined by the power of radio jets and the duty cycle of RLAGNs.
Previous research has shown that for massive, cluster-scale halos, the energy contained in observed X-ray cavities  \citep{McNamara07,Heckman14} compares reasonably well with the cooling rate as traced by the X-ray luminosity. 
Direct observational evidence for maintenance-mode feedback via RLAGNs exists for groups \citep{Werner12,Werner14}, but not for $L^{*}$ galaxies that occupy halos with masses of $\sim10^{12}\,M_{\odot}$.
Our finding that the RLAGN fraction is not dependent on the axis ratio of their host galaxies suggests that the maintenance-mode feedback is also important for low-mass quiescent galaxies with disc-like morphology.
To estimate whether or not the observed radio AGNs in $L^{*}$ galaxies (log ${ {M_{*}}/{ M_{\odot}}}\approx10.7-11$, or $\sigma_{*}\approx180-200 \,\rm km\,s^{-1}$) have a matching X-ray luminosity, we can either use the observed $L_{X}-M_{\rm halo}$ scaling relation $L_{X}\propto M_{\rm halo}^{1.65}$ {  of} \citet{Wang14} or the simplest theoretical expectation for the scaling of $L_{X}$ with $M_{\rm halo}$ based on the virial theorem ($L_{X}\propto M_{\rm halo}^{{4}/{3}}$).
The most massive halos (log ${  M_{\rm halo} /M_{\odot}})\approx 15$) host the most luminous radio galaxies ($L_{150\rm MHz}\approx 10^{25}-10^{26}\, W\,Hz^{-1}$), and the duty cycle of such a RLAGN can be inferred from the observed fraction of luminous RLAGNs at the high-mass end, assuming that all galaxies at a given mass go through recurrent RLAGN activity.
We assume that the conversion from radio power to kinetic power and the coupling of this kinetic power to the hot phase on average do not depend strongly on halo mass.
Therefore, the required RLAGN luminosity for $L^{*}$ galaxies must be of order $L_{150\rm MHz}\approx 10^{21}\, W\,Hz^{-1}$ in order to offset cooling if they have similar duty cycle to massive galaxies.
This luminosity requirement should be independent of the morphology of galaxies if maintenance-mode feedback is a general heating mechanism. 
In our work, $L_{150\rm MHz}= 10^{21}\, W\,Hz^{-1}$ happens to be the detection limit, and for $L^{*}$ galaxies the observed RLAGN fraction is $\sim10$\%, which is similar to the luminous ($L_{150\rm MHz}\gtrsim  10^{25}\, W\,Hz^{-1}$) RLAGN fraction at the very high-mass end \citep{Sabater19}. 
We conclude that the observed RLAGNs in $L^{*}$ quiescent galaxies are in principle sufficiently luminous to offset cooling and maintain the lack of star formation.
This is especially important for the oblate galaxies, because they typically do not host very powerful RLAGNs as revealed in  \citetalias{Barisic19} and this work.
We emphasize that this is only a crude estimate and only serves to illustrate that the low observed radio luminosity in $L^{*}$ quiescent galaxies cannot be used to argue against the importance of maintenance-mode feedback in low-mass halos.}
}
}

\section{Conclusion}\label{sec:conclusion}

In this work, we combined the LoTSS DR1 and SDSS DR7 data in the HETDEX region to form a sample of 15934 colour-based quiescent galaxies with 1912 RLAGNs brighter than \lradio$= 10^{21} \rm\,W\,Hz^{-1}$.
To remove possible biases from the shape--\mstar and \frl--\mstar dependence, we carefully divided the whole sample into groups according to their \mstar, $\sigma_{*}$, \mr and projected axis ratio $q$.
Directly plotting the RLAGN fraction \frl versus $q$ of different groups and comparing the RLAGN hosts with control samples provides some interesting information concerning the connection between RLAGNs and the shape of their hosts.
We list the main  conclusions that we make from this work as follows:
\begin{itemize}
    \item The \frl of most of the subsamples, especially the low-mass ones, does not show a simple positive correlation with the projected axis ratio $q$, which is found for high-luminosity RLAGNs by  \citet{Barisic19}. 
    Both {highly elongated} and round galaxies have a substantial fraction of RLAGNs. 
    \item High-luminosity RLAGNs are typically rounder than normal galaxies.
    \item The $q$ distributions of low- and intermediate-\lradio AGNs are not significantly different from the \mstar(or $\sigma_*$ ,\mr)-controlled sample.
    \item We fit the $q$ distributions to obtain the triaxial and oblate galaxy fractions for RLAGNs and all galaxies. Both T- and O-type sources have an increasing RLAGN fraction with \mstar, $\sigma_*$ and \mr.
    \item Although T-type sources have a higher RLAGN fraction generally, the difference between T- and O-type results is much less than that found by  \citet{Barisic19} because of the contribution of low-power RLAGNs in O-type galaxies.
    \item {We find that the $q$ distributions of the in-cluster RLAGNs and not-in-cluster control sample do not show a significant difference. Therefore, the morphology--RLAGN relation is not or only marginally influenced by the environment.}
    \item {T/O-type classification is not likely to have a simple match with the accretion mode dichotomy (LERG/HERG) according to the source numbers and stacked spectra.}
    
\end{itemize}
Our findings show that RLAGNs and the morphology of their hosts are correlated, and this correlation is radio-power- and/or mass-dependent.
High-power RLAGNs are more likely to be in a visually round galaxy and vice versa, whereas low-power sources do not show a preference in the optical shape.

The difference between high- and low-power RLAGNs implies that they could have different triggering mechanisms.
Long-term evolution processes can influence the triggering probability and power of radio jets.
While high-power RLAGNs can be triggered in galaxies with a merger-rich history, low-power sources could be located in `dead' spirals.

This work is based on the first data release of the LoTSS survey.
The limited sample volume and depth make it still difficult for us to investigate the role of environment in the RLAGN--morphology relation and in evolution across cosmic time.
With the advance of the LoTSS project, the whole northern sky will be covered and a deeper survey will be achieved.
With these new data, it will be easier to study the relation of RLAGNs, morphology, and environment more thoroughly.
Furthermore, the upcoming WEAVE--LOFAR survey  \citep{Smith16} will provide more detailed host galaxy properties, especially the dynamic features.
This will eventually give us the answer of how morphology is connected to RLAGN power and provide important clues about the triggering of powerful radio jets.

\begin{acknowledgements}
We thank an anonymous referee for useful suggestions that have improved our paper.
Data were provided by the LOFAR Surveys Key Science project (LSKSP; https://lofar-surveys.org/), from observations with the International LOFAR Telescope (ILT). 
LOFAR \citep{vanHaarlem13} is the Low Frequency Array designed and constructed by ASTRON. 
It has observing, data processing, and data storage facilities in several countries, that are owned by various parties (each with their own funding sources), and that are collectively operated by the ILT foundation under a joint scientific policy. 
The efforts of the LSKSP have benefited from funding from the European Research Council, NOVA, NWO, CNRS-INSU, the SURF Co-operative, the UK Science and Technology Funding Council and the J\"{u}lich Supercomputing Centre.
XCZ acknowledges support from the CSC (China Scholarship Council)-Leiden University joint scholarship program.
PNB is grateful for support from the UK STFC via grant ST/R000972/1
AvdW acknowledges funding through the H2020 ERC Consolidator Grant 683184.
MB acknowledges support from INAF under PRIN SKA/CTA FORECaST and from the Ministero degli Affari Esteri della Cooperazione Internazionale - Direzione Generale per la Promozione del Sistema Paese Progetto di Grande Rilevanza ZA18GR02.
HJAR and WLW acknowledges support from the ERC Advanced Investigator programme NewClusters 321271. 
WLW also acknowledges support from the CAS-NWO programme for radio astronomy with project number 629.001.024, which is financed by the Netherlands Organisation for Scientific Research (NWO).
IP acknowledges support from INAF under the SKA/CTA PRIN ``FORECaST" and the PRIN MAIN STREAM ``SAuROS" projects.
\end{acknowledgements}

%
%
\bibliographystyle{aa}
\bibliography{RLwithshape}
\begin{appendix} 
\section{Axis ratio versus redshift}\label{app:qz}
\begin{figure*}[ht]
\includegraphics[width=\linewidth]{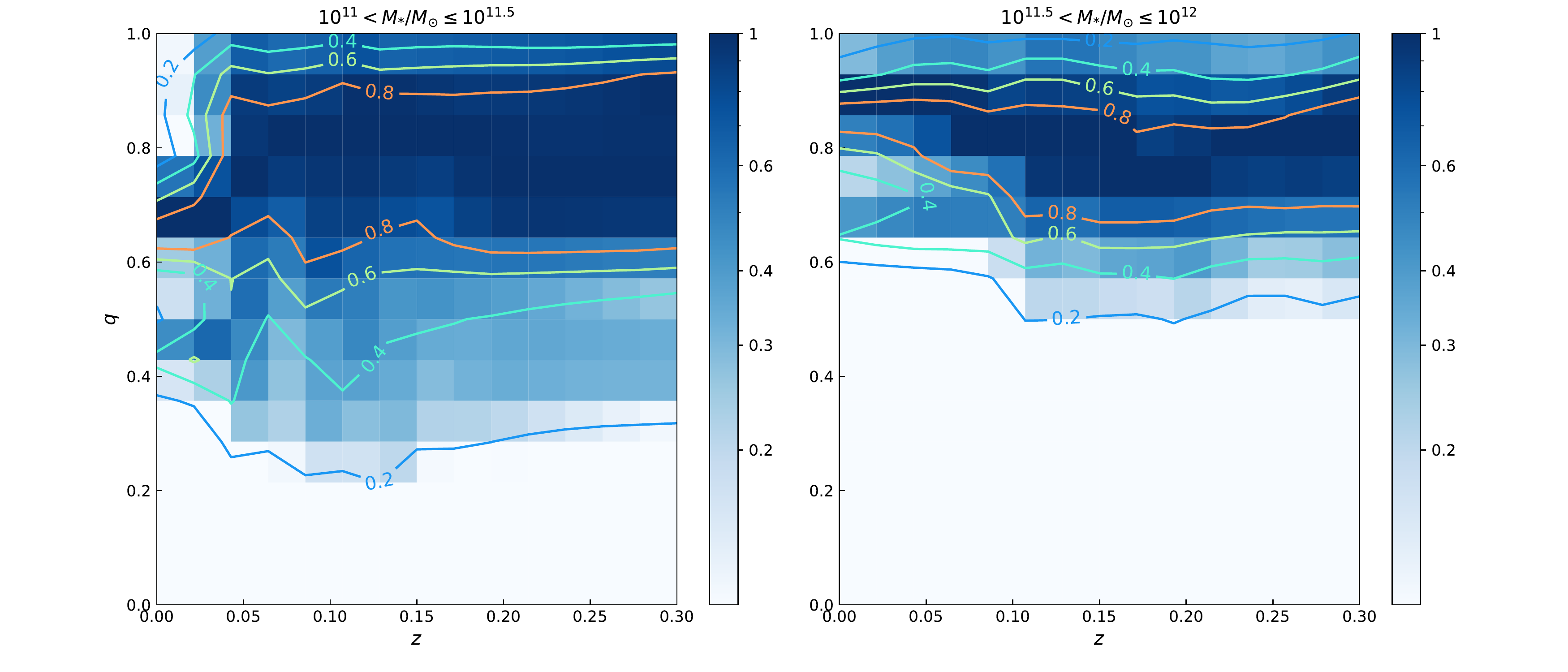}
\caption{Normalized distribution of galaxies at fixed \mstar in the $q-z$ plane. 
Kernel-density estimation was performed to reduce the error from binning.
The color of each cell represents the source number relative the the maximum source number cell at fixed $z$.}
\label{fig:qz}
\end{figure*}

{To investigate whether the observed axis ratios of high-redshift galaxies in our work are biased due to small angular sizes, we selected two massive galaxy samples from the 33324 galaxies in SDSS DR7 (see Sect. \ref{sec:data}) with log$($\mstar$/M_{\odot})\in(11,11.5]$ and $(11.5,12]$, because only massive galaxies can be detected at high redshift as shown by Fig. \ref{fig:mldist}. 
The evolution of the observed axis ratio distributions of the two sample is shown in Fig. \ref{fig:qz}. 

If the bias from smaller angular sizes is significant, the axis ratio should be larger at higher redshift.
However, we do not find significant variations in the distribution of $q$ at high redshift.
We conclude that the bias of axis ratio measurement due to smaller angular size of galaxies at high redshift is not important in this work.

} 
\end{appendix}

\end{document}